\documentclass[useAMS,usenatbib]{mn2e}

\usepackage{epsfig}
\usepackage{lscape} 

\input{psfig.sty}

\title[Galaxy Tomography]{Beating Lensing Cosmic Variance
with Galaxy Tomography }

\author[Ue-Li Pen]{
Ue-Li Pen,$^{1}$\thanks{E-mail:\ pen@cita.utoronto.ca}
\\
$^1$\,Canadian Institute for Theoretical Astrophysics, University of
Toronto, M5S 3H8, Canada 
}

\begin{document}


\pagerange{\pageref{firstpage}--\pageref{lastpage}} 
\pubyear{2004}

\maketitle

\label{firstpage}

\begin{abstract}

I discuss the use of cross correlations between galaxies with distance
information and projected weak lensing dark matter maps to obtain a
fully three dimensional dark matter map and power spectrum.  On large
scales $l\la 100$ one expects the galaxies to be biased, but not
stochastic.  I show that this allows a simultaneous solution of the
full 3-D evolving galaxy bias and the dark matter power spectrum
simultaneously.  Within the photo-z information of the CFH lensing
legacy survey, this allows a threefold reduction of statistical error,
while a cross correlation with CLAR or other deep spectroscopic surveys
allows a tenfold improvement in dark matter power accuracy on
linear scales.  This makes lensing surveys more sensitive to the
cosmic equation of state and the neutrino masses.

\end{abstract}


\newcommand{\be}{\begin{eqnarray}}
\newcommand{\ee}{\end{eqnarray}}
\newcommand{\beq}{\begin{equation}}
\newcommand{\eeq}{\end{equation}}

\section{Introduction}

The great success of modern physical cosmology has been driven by
precision measurements of the cosmic microwave background (CMB), whose
properties can be computed to high accuracy from first principles.
Current CMB fluctuations have been measured to high statistical
accuracy, which describes the universe at recombination.  To constrain
fundamental cosmological parameters, such as the matter density of the
universe, one requires measurements at low redshift as well.

The low redshift measurements are difficult.  Galaxy distributions can
be measured in three dimensions and to high accuracy, but their
formation is not understood in any detail.  This leaves a systematic
error in using their power spectrum to infer cosmological parameters.
While the CMB power spectrum has been measured to statistical
accuracies better than 1\%, galaxy power spectra are not reliable
tracers of matter at any accuracy close to that.  It is known that
galaxies of different morphologies are biased relative to each other.
To complicate matters further, it is now known that the bias depends
on scale, and that their distribution is stochastic on non-linear
scales \citep{2003astro.ph.10725T,2003ApJ...594...33F}.  This bias and
stochasticity is expected to evolve with redshift.  The interpretation
of their distribution is difficult to quantify, and galaxy power
spectra can only constrain fundamental parameters to limited
precision.  The WMAP team \citep{2003ApJS..148..175S} combined the CMB
power spectrum with the 2dF galaxy power spectrum.  For the latter,
the power is measured to high statistical accuracy, but with an
unknown bias, so its interpretation is limited by systematic errors
and was not used in their analysis.  The WMAP team instead used only
the slope of the galaxy power spectrum, which is hoped to be a less
biased indicator of the underlying dark matter distribution.  But it
is not independently known how large an error that might have, so the
error on the error bar is very difficult to quantify.  Using no
external data, WMAP data allows even closed cosmological models with
no cosmological constant.

In contrast, the distribution of dark matter can be predicted to high
accuracy through perturbation theory and simulations, since their
equations of motion are simple and can be modelled to high precision
with N-body simulations.  Its measurement is intrinsically very
challenging.  Statistical gravitational lensing of background galaxies
in blank fields has been a very successful approach.  But the signals
are intrinsically weak, limiting the accuracy of the constraints.
Nevertheless, the cleanness of the data has provided some of the
strongest low redshift cosmological constraints
\citep{2002A&A...393..369V,2003PhRvL..90v1303C}.

One intrinsic limitation of gravitational lensing studies is the broad
redshift distribution of lenses.  The observed shear signal integrates
over all the matter distribution along the line of sight.  Using
multiple source planes gives some handle on the distance, but even in
the best of cases only very crude redshift resolution can be
extracted\citep{2002PhRvD..66h3515H}.  Sample variance of lensing
surveys, and fundamentally the cosmic variance of an all sky survey,
are limited to the number of two dimensional modes on the sky.  If one
could beat this cosmic variance, one could significantly improve on
the results.

Galaxies, on the other hand, allow measurements of accurate redshifts.
One can measure their redshifts from optical or radio spectroscopy.
Their sample variance is smaller by the volume to surface area ratio
of a survey.  But as mentioned previously, the distribution of
galaxies is known to be biased, and this bias is likely to depend on
scale and cosmic time.  Their evolution is unlikely to provide a
direct measure of the cosmic equation of state.
\citet{2003PhRvL..91n1302J} have proposed a purely geometric procedure
to measure the geometry of the universe.  Our proposal differs by
actually solving for the matter distribution, which allows much higher
potential accuracy.

In this paper I show how one can combine the redshift resolved
distribution of galaxies to significantly improve the information of
lensing surveys.  It is possible to simultaneously measure the
evolution of bias as a function of scale and time, and thus the 3-D
distribution of dark matter.  The only intrinsic limit is
stochasticity between galaxies and dark matter.

\section{Strategy}

Perhaps this sounds like a free lunch.  Just counting degrees of
freedom, the evolution of bias as a function of scale and time is a
two dimensional function.  The observable dark matter power spectrum
is a one dimensional function of angular scale.  The cross
correlation, however, does have two dimensions as well.  The map data
set is also two dimensional on the sky, so the number of degrees of
freedom adds up that one could in principle measure the dark matter
evolution using galaxy cross correlations.

To illustrate the procedure, we will consider a simplified case where
the three dimensional quantities are projected along a Cartesian axis,
say $z$.  The analogy of this axis is distance along the line of
sight, which also corresponds to the direction along which we expect
cosmological evolution to be important.  One generically expects the
bias to depend on cosmic time.  If bias did not depend on the length
scale tangential to the line of sight, the procedure is
straightforward.  We assume that we know the three dimensional
distribution of galaxies up to an unknown bias function $b(z)$,
$\delta_g(x,y,z)=b(z)\delta(x,y,z)$.  Any spatial dependence of the
bias will be captured by the cross correlation
\citep{1998ApJ...504..601P} coefficient, which we will discuss later.
The two dimensional distribution of dark matter is projected by a
known lensing kernel $w_L(z)$, so the lensing surface density $\kappa$
is $\kappa(x,y)=\int \delta(x,y,z) w_L(z) dz$.  For this example, we
neglect the change of angular scales with distance.  We can form a
galaxy projection weighted by an unknown weight function $w(z)$.  For
practical purposes, this weight function will be smooth, so it
suffices to sample it at a number of points $n_z$.  The galaxy surface
density in terms of this yet to be determined function is
\begin{equation}
\Sigma_g(x,y)=\sum_{i=1}^{n_z} \delta_g(x,y,z_i)b(z_i)w(z_i) \Delta
z_i.
\label{eqn:sigma}
\end{equation}
We define a residual function $e=\int (\Sigma_g-\kappa)^2 dxdy$.  One
can then solve for the weight function by differentiating for weights
at redshift $z_i$.  This
results in a coupled system of equations
\begin{equation}
\int dxdy \delta_g(x,y,z_i) \Sigma_g(x,y)=\int dxdy \kappa(x,y)
\delta_g(x,y,z_i). 
\label{eqn:cpl}
\end{equation}
The unknowns $w_L(z_i)$ appear implicitly linearly in equation
(\ref{eqn:cpl}) through the surface density $\Sigma_g$ and equation
(\ref{eqn:sigma}). 
One can solve this system of equations to obtain
$w(z_i)=w_L(z_i)/b(z_i)$ if the intrinsic distributions of galaxies
and dark matter are the same, there is no noise, and the
number of pixels $n_{\rm pix}$ in the map is larger 
than the number of unknowns in the redshift dimension.  Typically, the
bias is a slowly evolving function of time. One
would bin in coarse intervals, for example a dozen photometric redshift 
bins.   The
image has many effective pixels.  With the exact solution, the
weighted galaxy map will be identical to the dark matter map.

In the presence of noise,
we can write Equation (\ref{eqn:cpl}) as a linear system
\begin{equation}
{\bf A} w=\kappa \pm \sigma
\label{eqn:a}
\end{equation}
where ${\bf A}$ is an $n_{\rm pix}$ by $n_z$ matrix.  We define a
noise matrix ${\bf N}$.  If the primary source of noise is shot noise in the
lensing map, this matrix is diagonal with entries being the variance
at each pixel $\sigma^2(x,y)$.  One solves the minimum variance solution
to (\ref{eqn:a}) as
\begin{equation}
w=({\bf A}^t {\bf N^{-1} A} )^{-1} {\bf A}^t {\bf N^{-1}} \kappa.
\label{eqn:w}
\end{equation}
In general one has $n_{\rm pix} \gg n_z$, so the inverse operation
is well defined.

One can in a similar fashion compute scale dependent bias evolution.
We define the scale dependent bias as the ratio of the power spectrum
of galaxies to that of the dark matter,
\begin{equation}
b^2(k,z)=\frac{P_{\rm gal}(k,z)}{P_{\rm dm}(k,z)}.
\end{equation}
The power spectrum is a two point statistic, which depends on the
square of the bias.
In terms of the density distribution, this relates the Fourier
transform of the galaxy density field linearly through $b(k,z)$ to
that of the dark matter:
\begin{equation}
\delta_g(\vec{k},z)=b(|k|,z)\delta(\vec{k},z),
\label{eqn:dgrz}
\end{equation}
or equivalently states that the galaxy density field is a convolution
over the dark matter distribution
$\delta_g(\vec{x},z)=\int \delta(\vec{x},z) b(|x|,z) d^3x$.
A bit of care needs to be taken in the interpretation of this Fourier
transform.  We have represented the three spatial coordinates
$(x,y,z)$ by a vector $\vec{x}$.  The radial direction $z$ is now the
third component of the $\vec{x}$.
The three dimensional expression in equation
(\ref{eqn:dgrz}) assumes that $b(z,r)$ varies sufficiently slowly as a
function of $z$ that one can apply a transform at each interval
$z_i$.  This is reasonably accurate assumption except for the largest
scale, where one would want to solve a general linear parametrized
system analogous to (\ref{eqn:w}) instead of Fourier transforming.

The density can be thought of as a
one-point distribution, which depends linearly on the bias.
For this redshift $z$ and scale $k$ dependent bias $b(z,k)$ we can
Fourier transform from wave number space $k$ to separation $r$:
\begin{equation}
b(z,r)=\frac{1}{(2\pi)^2}\int b(z,k) \frac{\sin(kr)}{kr} k^2 dk.
\label{eqn:bz}
\end{equation}

Our redshift binned bias function (\ref{eqn:sigma}) now generalizes to
a two dimensional function of redshift and separation.  We then
convolve each redshift slice by the parametrized $b(z,r)$.  We then
solve for the spatial and temporal structure of the bias.  In
practice, the bias should evolve slowly in space and time.  Once we
know the evolution of bias as a function of length scale and cosmic
time, it is straightforward to obtain the dark matter power spectrum
as a function of scale and time.  In the linear regime, this measures
the growth factor.  In the non-linear regime, the PD96
\citep{1996MNRAS.280L..19P} approach gives a mapping between linear
and non-linear power spectra.  This needs to be calibrated against
simulations, but as we see below, the inferences are probably
systematically limited by other non-linear effects.  The growth factor
in turn maps directly onto the equation of state of space-time and
matter, i.e. dark energy and neutrino mass
\citep{2002PhRvD..66h3515H}.

\section{Limitations}

The improvement is intrinsically limited on several fronts.  It requires
that bias be linear and non-stochastic \citep{1998ApJ...504..601P}.
On non-linear scales, the power spectrum of galaxies is known to
be stochastic \citep{2003MNRAS.346..994P,2002ApJ...577..604H}.  In the cross
correlation tomography, one can measure the degree of stochasticity
which provides a fundamental limit to the accuracy.  Other sources of
limitations are the noise in the lensing and galaxy power measurement.

One can measure the impact of  stochasticity.  We use the model
from \citet{2002ApJ...577..604H}.  The power spectrum of galaxies is
related to the dark matter by the bias, $P_g=b^2 P$, and the cross
correlation coefficient $r(k,z)$ can depend on time and scale.  It
relates $\langle\delta_g \delta \rangle=r \sqrt{P_g P}$.  At the two
point level, the bias and cross correlation completely describe all
two point statistics, and parametrizes all effects including
non-linearity.

In the presence of stochasticity, the solution using our procedure becomes 
\begin{equation}
w(z_i)=\frac{w_L(z_i)r(z_i)}{b(z_i)}.
\label{eqn:s}
\end{equation}
The problem is that one needs to know the cross correlation parameter $r$
to infer the true dark matter power spectrum.  In the previous section
we implicitly assumed $r=1$, which is expected on linear scales.

While we cannot simultaneously measure the stochasticity and bias, we
can infer an integrated upper bound.  Once we have solved for the
weight function $w$, we define a residual $\chi^2$ as
\begin{equation}
\chi={\bf S}^{-1}({\bf A} w-\kappa)
\end{equation}
In the absence of noise, this is the lensing power which was not
predicted from the galaxies, and corresponds to the integral of
$(1-r(z)^2)w_L^2(z)$ over the lensing weight.  If this integral is small, it
places an upper bound on the variation of $r$ at {\it any} redshift.

With this framework in mind, we can substitute the actual parameters.
We will use a linear power spectrum evolution model,
$\Delta^2(z)=\Delta^2(z=0)D(z)^2/(1+z)^2$.  The notation $\Delta^2
\equiv k^3 P(k)/2\pi^2$ is taken from
\citet{1999coph.book.....P}.

The comoving angular diameter distance is
\begin{equation}
\chi(z)=c\int_0^z {\frac{dz}{H(z)}}
\end{equation}
where H(z) is the Hubble constant at redshift z:
\begin{equation}
H(z)=H_0[(1+z)^2(\Omega_m z+1)-\Omega_{\Lambda}z(z+2)]^{1/2}.
\end{equation}
For the angular diameter distance $\chi$ we use the fitting formula from
\citet{1999ApJS..120...49P}.

For dark matter the lensing weight is
\begin{equation}
w_L(z)=\frac{3}{2}\Omega_m {H_0}^2g(z)(1+z)
\end{equation}
where
\begin{equation}
g(z)=\chi(z)\int_z^{+\infty}dz'n_s(z')\frac{\chi(z')-\chi(z)}{\chi(z')}.
\end{equation}
$n_s(z)$ is the normalized distribution of source galaxies. The two dimensional
dark matter map is then
\begin{equation}
\kappa(\theta_x,\theta_y)=\int \delta(d_A(z) \theta_x,d_A(z)\theta_y) w_L(z)
      \frac{d\chi}{dz} dz.
\end{equation}
Analogously, we define a weighted galaxy surface density with respect
to a yet undetermined weight $w(z)$
\begin{equation}
\Sigma_g(\theta_x,\theta_y)=\int
\delta_g(d_A(z)\theta_x,d_A(z)\theta_y,z)
w(z_i) d z.
\end{equation}
In analogy to equation (\ref{eqn:w}) we solve for $w(z)$ and infer the
combination of bias and stochasticity given by equation (\ref{eqn:s}).

\section{Applications}

We consider parameters for the CFHT Legacy Survey
\citep{2003astro.ph..5089V}, and its expected accuracy.  We use a
source density comparable to the VIRMOS-DESCARTES survey of 30
galaxies per square arcminutes\ with a median source redshift of
$z_0=1$.  The mean ellipticity is taken to be $\epsilon=0.3$.  The
survey area is 200 square degrees, which is half a percent of the sky.
We neglect the Poisson noise from the cross correlation lens plane
galaxies, which is much smaller than the lensing noise in each
redshift slice.

The LS lensing survey measures the projected two dimensional dark
matter distribution.  In principle, the three dimensional power
spectrum estimation requires inverting the Limber equation, which can
be numerically unstable.  We will use the half weighted approach of
\citet{2003MNRAS.346..994P}, which leads to mostly uncorrelated error
bins in the three dimensional power spectrum.  For simplicity, we use
the power spectrum at the angular diameter distance of the typical source
redshift $z_0=0.4$.  The mapping becomes $k=l/\chi(z_0)\sim (l/1000) h^{-1}$
Mpc$^{-1}$.  We bin the power spectrum in factors of 2 in $l$, and use
Gaussian errors $\Delta C_l=(C_l+C_l^N)/\sqrt{(2l+1)\Delta l f_{\rm
sky}}$.

The assumption of zero stochasticity is likely to break down when the
galaxy clustering is non-linear.  We model this conservatively as the
stochasticity parameter being inversely proportionate to the variance
\begin{equation}
r\sim \frac{1}{1+\Delta^2}.  
\label{eqn:r}
\end{equation}
Current data suggests that galaxies and dark
matter are probably better correlated than that
\citep{2003MNRAS.346..994P,2002ApJ...577..604H}, so tomography will 
likely work better than this estimate.  The good news here is that
one will be able to measure the stochasticity, and know the reliability
of the inference.

\begin{figure}
\vskip 3.3 truein
\includegraphics{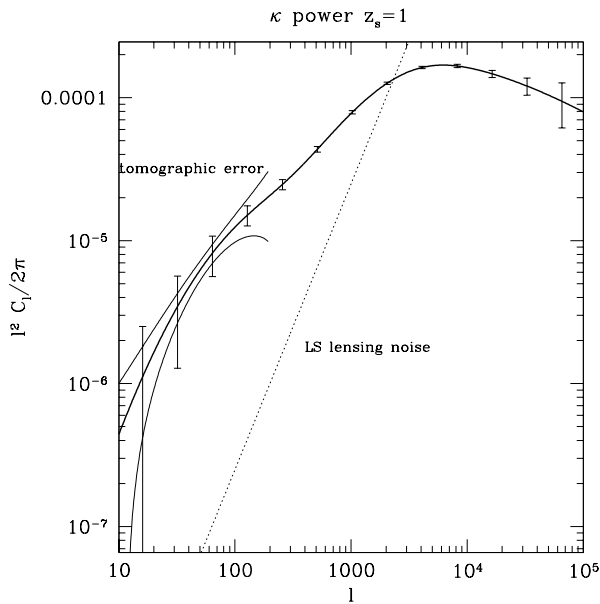}
\caption{
The expected lensing power spectrum for the CFH Legacy Survey.  Error
bars show the variance due to lensing.  The upper and lower thin lines
show the variance if galaxy tomography is used.  In our model, we assume
stochasticity proportionate to non-linearity, which limits the tomography.
Realistic stochasticities are smaller, so the procedure may work to
smaller angular scales (larger l).  The dotted line is the lensing shot noise
power spectrum.
}
\label{fig:cfhtls}
\end{figure}

Figure \ref{fig:cfhtls} shows the expected accuracy of the LS
dark matter power spectrum measurement.  To obtain the level of
stochasticity in equation (\ref{eqn:r}), we took the matter power
spectrum at $z=0.4$.  We assumed $l/4$ radial modes along the line of
sight.  This is a rough estimate, based on a comoving radial size of
$0.5 h^{-1}$ Gpc for the reconstructed galaxy distribution.
The two solid lines show the accuracy with a matched redshift survey,
for example CLAR \citep{2000SPIE.4015...33C}. If only photometric
redshifts are available, the improvement can be limited by the square root
of the number of radial redshift bins, if these are less than the
number of radial modes.  At $l=50$, we need 12 radial bins to a
redshift of 1, which may be achievable with photometric redshifts.
With a lower fidelity, the errors will grow.

There are other sources of errors which we have not included in the
figure.  There are inevitably errors on $b(z,r)$.  When stochasticity
is small, one would in general expect the error on the bias to the
smaller than sample variance.  If the dark matter and galaxy maps
differ by a multiplicative constant, one can just divide the maps and
measure this constant perfectly, with no sampling variance.  
Measurement noise does limit at some level how well this can be done,
which depends on both the lensing noise and the galaxy distribution noise.

We see that we obtain significant improvement in the linear scale
power spectrum estimation.  The total statistical information is still
dominated by the smaller non-linear scale power, for which many more
modes can be measured.  These non-linear scales in principle encode
the power spectrum and its history, but require detailed N-body
simulations to calibrate.  There are also questions of fundamental
limits to the precision of non-linear scale power, since baryons
back react on the dark matter.  The distribution of baryons is
intrinsically harder to predict since they can heat and cool in ways
that is not predictable from first principles.

Future lensing surveys in the next five years can cover the whole sky,
and can map sources at distances beyond the epoch of reionization at
$z>7$ \citep{2003astro.ph..5387P}.  It is also possible to map the
redshift distribution of galaxies to similar distances on the whole
sky with the Square Kilometer Array.  This allows one to beat the
normal sample variance limitation, also sometimes called 'cosmic
variance'.  In an interval of angular wavenumber $l-2l$ there are
$\sim 3l^2$ two dimensional modes on the sky.  This can in general be
increased to $\sim l^3$ modes using tomography.  But actual accuracy
that can be gained will depend on the effects of non-linearity and
stochasticity in the distribution of galaxies.

\section{Conclusions}

Gravitational lensing is a clean procedure to measure the power
spectrum of dark matter at low redshifts.  It is already providing
competitive constraints on cosmological parameters, and does not make
any assumptions about the nature of galaxy biasing nor any other
gastrophysical process.

Like the cosmic microwave background, gravitational lensing accuracy
is limited by sample variance.  It can only provide two dimensional
power spectra.  Some of the most important open cosmological questions
involve precision measurements of the cosmic equation of state.  For
this procedure to be successful, one needs  accurate measurements of
the dark matter power spectra, ideally in the linear regime and at
different redshifts.  This is a regime where gravitational lensing is
significantly restricted in accuracy.

We have shown in this paper how one can improve on the accuracy by
using the cross correlation to galaxies with redshift information.
One can reduce the errors without any assumptions on
galaxy bias, even if it evolves and depends on scale.  The only
requirement is a low stochasticity.  The total actual stochasticity of
the data set is measurable, and can be folded into the error
analysis.  

We estimated this improvement on two concrete surveys: the
Canada-France-Hawaii-Telescope Legacy Survey Weak Lensing Survey, and
the Canadian Large Adaptive Reflector deep survey.  With the added
distance information, one can reduce the error bars on linear scales
$l\sim 50$ by up to a factor of 3.

I would like to thank Uros Seljak for helpful discussions.

\bibliography{penbib}
\bibliographystyle{mn2e}

\appendix

\bsp

\label{lastpage}

\end{document}